\documentstyle[prc,aps,epsfig]{revtex}
\begin{document}
\tightenlines
\title
{Rho properties in a hot meson gas}

\author
{Ralf Rapp$^{1}$ and Charles Gale$^{2}$}

\address
{1) Department of Physics and Astronomy, State University of New York,
    Stony Brook, NY 11794-3800, U.S.A.\\
 2) Physics Department, McGill University, Montreal, Quebec H3A 2T8,
    Canada \\}

\maketitle

\begin{abstract}
Using effective meson Lagrangians we study the interaction of rho
mesons in a hot baryon-free system.  Various mesonic resonances 
in direct $s$-channel reactions are investigated employing standard 
self-energy techniques, including new reactions that
have up to now not been considered in a self-consistent approach
at finite temperature. The importance of 
subthreshold resonances, which are readily accounted for  
through off-shell effects within our framework, is emphasized.    
Special care is taken in reproducing radiative decay widths, as they   
provide valuable  constraints on the evaluation of dilepton spectra. 
In particular, we compare our results for dilepton production  
rates to earlier calculations based on an incoherent summation of 
individual processes.  
\end{abstract}


\section{Introduction}
\label{intro}
Dilepton measurements in heavy-ion collisions at
intermediate~\cite{DLS} and high~\cite{CERES,HELIOS3} bombarding
energies have revealed a strong excess of pairs as compared to
proton-induced reactions. It has become clear that the collection
of hadronic sources that can successfully account for the measured
spectrum in proton-induced reactions fails to reproduce the measured 
yield in nucleus-nucleus collisions. Even after the inclusion of (free)
$\pi^+\pi^-$ annihilation during the interaction phase of the
hadronic fireball the enhancement, especially for invariant
dilepton masses $M_{ll}$ below the $\rho$-mass, remained
unexplained \cite{drees98}. Thus many theoretical efforts have
concentrated on the role of medium effects in pion and rho meson
propagation. It seems fair to say that there currently exists two
main schools of thought that have staked a claim onto the theoretical
interpretation of the CERN low-mass dilepton measurements. The first
one assigns the rho meson mass as an order parameter of the chiral 
symmetry restoration. The CERN data, in this approach, 
then signals a dropping of the rho meson mass~\cite{BR91,CEK,LKB}. A 
second interpretation relies on the fact that
in a strongly interacting medium, the rho meson will have its
width greatly increased due to modifications of its pion 
cloud~\cite{AKLQ,HFN93,CS93,CRW,KKW97,UBRW}
 as well as to the direct coupling to baryonic 
resonances~\cite{FP97,RCW,RUBW,PPLLM}. Whether these two different scenarios 
can be reconciled, or whether one of them can be eventually ruled out,  
is the subject of much current research and debates.
In this article we follow the approach  that is
germane to the second of those theoretical avenues. We will restrict 
ourselves to a heat bath of pions, kaons and rho mesons, characterized 
by a finite temperature $T$. Using the many-body
formalism of Ref.~\cite{RCW}, we will separately investigate
each meson channel and calibrate its individual strength through 
empirical information on both hadronic and electromagnetic branching 
ratios, thereby 
introducing  some new channels that have not been considered before in a
framework like the one at hand.

The baryon-rich nuclear medium has been shown to strongly affect the
vector meson properties, see, 
{\it e.g.}, Refs.~\cite{RCW,FP97,KKW97,PPLLM,EIK98,gao98}. Then, a
self-consistent finite-temperature assessment of the meson
contributions is necessary to quantify this assertion. This is 
particularly true for CERN-SpS energies: in spite of the fact that 
the phase space is dominated by mesons (with a final pion-to-nucleon 
ratio of about 5:1), the explanation of the dilepton data in both 
the dropping mass and the in-medium broadening scenario crucially  
depends on baryonic contributions. At RHIC energies, however, it 
is hard to imagine that baryons could play a major role; since 
dilepton measurements in the PHENIX experiment will be able to 
address the invariant mass  region between $\omega(782)$ and $\phi(1020)$ 
with excellent resolution, in-medium properties of the rho in mesonic
matter should  be most relevant there.   
Various analyses in this direction have already been 
performed~\cite{GK91,gali,ha95,song,SYZ1,RCW,GG98}. 
The moderate collisional broadening for on-shell  rho-mesons  
found in Ref.~\cite{ha95} was mainly attributed to resonant scattering
via intermediate $a_1(260)$ and $K_1(1270)$ states. 
Similar results were obtained 
in Ref.~\cite{RCW} using finite temperature self-energies which  
include off-shell effects.   
This is to be expected, since the $a_1(1260)$ and
$K_1(1270)$ are located above the free $\rho\pi$ and $\rho K$
thresholds, respectively, which makes the resonant contribution dominant. 
In order to properly address the low mass region, additional 
mesons have to be considered. Those larger meson ensembles have in fact
been included in calculations where individual rates are summed 
over~\cite{gali}. There, the contribution of the $\omega$, for example,
appears as a radiative decay channel. In the language used in this work,
the $\omega$ is a 
sub-threshold state, {\it i.e.}, in $\rho\pi\to\omega(782)$,
where, given a typical thermal pion energy of 300-400~MeV, the
relevant $\rho$-mass to form the $\omega$ would be substantially
off-shell, $M\simeq 400$~MeV.

The objective of this article is twofold: on the
one hand, we would like to examine if a consistent off-shell
treatment of the most important mesonic resonances has a more
severe impact on the dilepton production rates than what has been 
estimated so far within different frameworks. At
the same time, we can investigate possible interference (or
collectivity) effects in the  coherent summation of the various
self-energy contributions. Since many-body calculations of this
nature can be rather convoluted, we will exhibit an explicit  
channel decomposition of the considered resonances $R=\omega, h_1,
a_1, K_1, \pi'$ and $f_1$. An upper mass limit of 1.3 GeV enables us
to adequately address the phenomenology that concerns us in this
work. 
As the simplest, but by no means negligible, 
 finite temperature effect
in the pion cloud of the rho we will furthermore include the Bose-Einstein
enhancement in the $\rho\to \pi\pi$ decay width, which is not 
new~\cite{HK87,GK91}. 
 Our paper is organized as follows: the
next section  introduces the hadronic Lagrangians used in this work, 
 followed by the determination of the parameters in our model in 
sect.~\ref{sec_param}.
We then evaluate the resulting in-medium rho properties 
(sect.~\ref{sec_rhoprop}) and 
their impact on dilepton production
rates (sect.~\ref{sec_dilep}). Sect.~\ref{sec_concl} 
contains a summary and concluding remarks.

\section{Interaction Lagrangians}
\label{sec_lagr}
Our starting point is the model for the $\rho$-meson in free space
employed previously in Refs.~\cite{CS93,CRW,RCW}. 
Based on the standard $\rho\pi\pi$ interaction vertex (isospin structure 
suppressed), 
\begin{equation}
{\cal L}_{\rho\pi\pi} = g_{\rho\pi\pi} \ \pi \ p^\mu \pi
\rho_\mu \ ,
\label{Lrhopipi}
\end{equation} 
($p^\mu$: pion momentum) the bare $\rho$-meson
of mass $m_\rho^{0}$ is renormalized through the two-pion loop
including a once subtracted dispersion relation, 
giving rise to the vacuum self-energy
\begin{eqnarray}
\Sigma_{\rho\pi\pi}^0(M) & = & \bar{\Sigma}_{\rho\pi\pi}^0(M)
-\bar{\Sigma}_{\rho\pi\pi}^0(0) \ ,
 \nonumber\\
\bar{\Sigma}_{\rho\pi\pi}^0(M) & = & \int \frac{p^2 dp}{(2\pi)^2} \
v_{\rho\pi\pi}(p)^2 \ G_{\pi\pi}^0(M,p)  \  ,
\label{sigrho0}
\end{eqnarray}
with the vacuum two-pion propagator
\begin{equation}
G_{\pi\pi}^0(M,p)=\frac{1}{\omega_\pi(p)} \
\frac{1}{M^2-(2\omega_\pi(p))^2+i\eta}
 \ ; \quad \omega_\pi(p)=\sqrt{m_\pi^2+p^2} \
\label{Gpipi}
\end{equation}
and vertex functions 
\begin{equation}
v_{\rho\pi\pi}(p) 
= \sqrt{\frac{2}{3}} \ g_{\rho\pi\pi} \ 2p \ F_{\rho\pi\pi}(p) \ 
\label{vrhopipi}
\end{equation}
involving a hadronic (dipole) form factor $F_{\rho\pi\pi}$~\cite{RCW}
(cf.~also Eq.~(\ref{ff}) below). 
Resumming the two-pion loops in a Dyson equation gives the free $\rho$
propagator
\begin{equation}
D_\rho^0(M)=[M^2-(m_\rho^{0})^2-\Sigma_{\rho\pi\pi}^0(M)]^{-1} \ ,
\end{equation}
which agrees well with the measured $p$-wave $\pi\pi$ phase shifts 
and the pion electromagnetic form factor obtained within the vector 
dominance model (VDM).

To calculate medium corrections to the $\rho$ self-energy in a hot meson
gas, we will assume that the interactions are dominated
by $s$-channel resonance formation. At moderate temperatures relevant
for the hadronic gas phase, the light pseudoscalar Goldstone bosons
$P=\pi,K$ are the most abundant species.
We can group the various resonances in $\rho P$ collisions in
two major categories, namely vector mesons $V$ and
axial-vector mesons $A$. For the latter, a simple interaction
Lagrangian, compatible with chiral symmetry and electromagnetic current
conservation, is given by
\begin{equation}
{\cal L}_{\rho P A}=G_{\rho PA} \ A_\mu \ (g^{\mu\nu} \ q_\alpha
p^\alpha - q^\mu p^\nu) \ \rho_\nu \ P \ ,
\label{LrhoPA}
\end{equation}
although other choices are possible~\cite{GG98}. $\rho P$ scattering via
intermediate vector mesons $V$ is determined by  Wess-Zumino
anomaly terms, which are of unnatural parity and involve 
the four dimensional antisymmetric Levi-Civita tensor
$\epsilon^{\mu\nu\sigma\tau}$:
\begin{equation}
{\cal L}_{\rho PV}=G_{\rho PV} \ \epsilon_{\mu\nu\sigma\tau} \ k^\mu
\ V^\nu \ q^\sigma \rho^\tau \ P \ .
\label{LrhoPV}
\end{equation}
In both Lagrangians~(\ref{LrhoPA}) and~(\ref{LrhoPV}),  $p^\mu, q^\mu$ and
$k^\mu$ denote the four-momenta of the pseudoscalar, rho- and
(axial-) vector-mesons, respectively. As a third possibility 
$\rho P$ scattering can proceed via a pseudoscalar resonance; here 
we restrict ourselves to the process $\rho\pi\to\pi'(1300)$, which 
can be described by 
\begin{equation}
{\cal L}_{\rho P P'}=G_{\rho PP'} \ P' \ ( k \cdot q \ p_\mu - 
p \cdot q \ k_\mu) \ \rho^\mu \ P \ .  
\label{LrhoPP}
\end{equation}

In addition to that, we will
need a $\rho VA$ interaction vertex, which is also related to anomaly
terms~\cite{KM90}. We choose the following form,
\begin{equation}
{\cal L}_{\rho VA}=G_{\rho VA} \ \epsilon_{\mu\nu\sigma\tau} \
\ p^\mu V^\nu \ \rho^{\sigma\alpha} \ k_\alpha A^\tau
- \frac{\lambda}{2} \ (k_\beta A^\beta)^2 \ ,
\label{LrhoVA}
\end{equation}
which again satisfies the appropriate conservation laws.
$\rho^{\sigma\alpha}=q^\sigma \rho^\alpha-q^\alpha\rho^\sigma$ is
the usual field strength tensor. We have explicitly
written here the kinetic energy term of the axial vector field
where the constant $\lambda$ represents a gauge freedom connected 
with the axial-vector field~\cite{ItZu}. In what follows we take 
$\lambda=1$; this choice will be further motivated   
in sect.~\ref{sec_rhoprop}.

An important hint on the importance of resonances in dilepton production
is provided by their radiative
decay width, which constitutes the $M^2=0$ (photon) limit
of the timelike dilepton regime. In Table~\ref{tabres} we have collected
mesonic resonances which are accessible via $\rho$-induced excitations
and exhibit substantial decay rates into final states involving
either photons or rho-mesons (or both). We expect these to be the relevant
contributions to dilepton production. In particular, note that the 
$f_1(1285)$ has a large  radiative decay width of 1.65~MeV for 
$f_1\to\rho\gamma$, which led us to the consideration of the 
$\rho\rho f_1$ interaction, Eq.~(\ref{LrhoVA}).

As a first step one has to assign realistic coupling constants
in the effective Lagrangians.
They are adjusted to the experimental branching ratios of the resonances
into, if available, both $\rho h$ and $\gamma h$ ($h=\pi,K,\rho$). 
For reliable estimates
of the $\rho P$ decay widths it is important to include the finite width
of the $\rho$ (in particular for sub-threshold states like the 
$\omega(782)$ or $f_1(1285)$).
This is accomplished  by folding the expression for the width
at given $\rho$-mass $M$ with the $\rho$ spectral function
$A_\rho^0(M)=-2 {\rm Im} D_\rho^0(M)$.
For the axial-vector meson resonances in $\rho P$ scattering the vertex
of Eq.~(\ref{LrhoPA}) leads to
\begin{eqnarray}
\Gamma_{A\to \rho P}(s) & = & \frac{G_{\rho PA}^2}{8\pi s} \
\frac{IF \ (2I_\rho+1)}{(2I_A+1)(2J_A+1)}
\int\limits_{2m_\pi}^{M^{max}} \frac{M dM}{\pi} \  A^0_\rho(M) \ q_{cm}
\nonumber\\
 & & \quad\quad \times \ [\frac{1}{2}(s-M^2-m_P^2)^2
+M^2 \omega_P(q_{cm})^2] \ F_{\rho PA}(q_{cm})^2 \ ,
\label{gammaA}
\end{eqnarray}
and from Eq.~(\ref{LrhoPV}) one obtains for vector resonances
\begin{equation}
\Gamma_{V\to \rho P}(s)  =  \frac{G_{\rho PV}^2}{8\pi} \
\frac{IF \ (2I_\rho+1)}{(2I_V+1)(2J_V+1)}
\int\limits_{2m_\pi}^{M^{max}} \frac{M dM}{\pi} \  A^0_\rho(M)
\ 2 q_{cm}^3 \ F_{\rho PV}(q_{cm})^2 \ ,
\label{gammaV}
\end{equation}
with $q_{cm}$ being the three-momentum of the decay products in the 
resonance rest frame, $\omega_P(q_{cm})^2=m_P^2+q_{cm}^2$.
$IF$ is an isospin factor, $M^{max}=\sqrt{s}-m_P$, and $F_{\rho PR}$
($R=A,V$) are hadronic form factors that reflect the finite size of the 
fields that appear in the 
effective vertices. We take them to be of dipole form,
\begin{equation}
F_{\rho PR}(q_{cm})=\left(\frac{2\Lambda^2_{\rho P}+m_R^2}
{2\Lambda^2_{\rho P}
+\left[\omega_\rho(q_{cm})+\omega_P(q_{cm})\right]^2}\right)^2
\ ,
\label{ff}
\end{equation}
normalized to 1 at the resonance mass $m_R$. 
With the Lagrangian for the pseudoscalar resonance $P'=\pi'(1300)$, 
Eq.~(\ref{LrhoPP}), one arrives at 
\begin{eqnarray}
\Gamma_{\pi'\to \rho\pi}(s)  =  \frac{G_{\rho\pi\pi'}^2}{8\pi} \
\frac{IF \ (2I_\rho+1)}{(2I_{\pi'}+1)(2J_{\pi'}+1)}
\int\limits_{2m_\pi}^{M^{max}} \frac{M dM}{\pi} \  A^0_\rho(M)
 \ q_{cm}^3 \ M^2 \ F_{\rho\pi\pi'}(q_{cm})^2 \ .  
\label{gammaP}
\end{eqnarray}
For the $\rho VA$ vertex,
which in our case corresponds to the $f_1(1285)\to \rho\rho$ decay,
the spectral functions of both outgoing $\rho$ mesons have to be
integrated over. One has
\begin{eqnarray}
\Gamma_{f_1\to \rho\rho}(s) & = & \frac{G_{\rho\rho f_1}^2}{8\pi} \
\frac{n \ IF \ (2I_\rho+1)}{(2I_{f_1}+1)(2J_{f_1}+1)}
\int\limits_{2m_\pi}^{M_1^{max}} \frac{M_1 dM_1}{\pi} \  A^0_\rho(M_1)
\int\limits_{2m_\pi}^{M_2^{max}} \frac{M_2 dM_2}{\pi} \  A^0_\rho(M_2)
\nonumber\\
 & & \qquad\qquad\quad\times \ 2 s q_{cm}^3 \ F_{\rho\rho f_1}(q_{cm})^2 \ ,
\label{gammaf1}
\end{eqnarray}
where the factor $n=1/2$ ensures the proper symmetrization of the two
$\rho$ mesons in the final state.  The integration limits are
$M_1^{max}=\sqrt{s}-2m_\pi$, as before, and $M_2^{max}=\sqrt{s}-M_1$.

For each vertex one is left with two unknowns: the
coupling constant $G$ and the form factor cutoff $\Lambda_\rho$,
which should lie in some reasonable region for the hadronic scales at
hand, typically $\Lambda_\rho\le 1-2$~GeV. The hadronic decay 
widths are in fact  
not very sensitive to $\Lambda_\rho$, since the dominant contributions
in the $M$-integrals are centered around the $\rho$-peak in
$A^0_\rho(M)$ (or towards the maximal $M$ if $m_R < m_\rho$), where
the three-momenta are rather small. More stringent constraints on 
$\Lambda_\rho$ are imposed by the radiative decay widths, since the
massless photon can carry away the maximal three-momentum. Invoking
the phenomenologically well-established (especially for purely mesonic 
processes) vector dominance model (VDM), the photon decay widths
follow from the hadronic couplings to vector mesons by simply (i) 
taking the $M^2\to 0$ limit, {\it i.e.}, substituting
 $A^0_\rho(M)=2\pi \delta(M^2)$ for real photons,
(ii) supplying the VDM coupling constant $(e/g)^2\simeq 0.052^2$,
and, (iii) omitting the $(2I_\rho+1)$ isospin degeneracy factor for
the final state. This yields for both axial-/vector resonances 
($R=A,V$)
\begin{equation}
\Gamma_{R\to \gamma P} = \frac{G_{\rho P R}^2}{8\pi} \
\left(\frac{e}{g}\right)^2 \ \frac{IF}{(2I_{R}+1)(2J_{R}+1)} \
2q_{cm}^3 \ F_{\rho P R}(q_{cm})^2 \ ,
\label{gammaph}
\end{equation}
whereas for the $f_1\to \gamma\rho$ decay one still has to integrate
over one $\rho$ mass distribution,
\begin{equation}
\Gamma_{f_1\to \gamma \rho} = \frac{G_{\rho\rho f_1}^2}{8\pi} \
\left(\frac{e}{g}\right)^2 \ \frac{2 n IF}{(2I_{f_1}+1)(2J_{f_1}+1)} \
\int\limits_{2m_\pi}^{\sqrt{s}} \frac{M_2 dM_2}{\pi} \  A^0_\rho(M_2)
\ 2 s q_{cm}^3 \ F_{\rho\rho f_1}(q_{cm})^2 \ .
\label{gammaf1ph}
\end{equation}
The additional factor of 2 accounts for the two possibilities of
attaching the photon to either of the outgoing $\rho$'s. Within the 
simple version of VDM employed here, the radiative decay of the $\pi'(1300)$ 
vanishes.

\section{Determination of Free Parameters}
\label{sec_param}
Let us now discuss the individual resonances, based on the
interaction vertices formulated above, in more detail. 

 The $\rho\pi a_1(1260)$ coupling constant has been estimated in 
our framework in Ref.~\cite{RCW} as $G_{\rho\pi a_1}=
13.20$~GeV$^{-1}$ taking~\cite{PDG} $\Gamma_{a_1\to\pi\rho}=400$~MeV,
$m_{a_1}=1230$~MeV and assuming a cutoff $\Lambda_{\rho\pi
a_1}=2$~GeV. With these parameters the radiative decay width of
the $a_1$ turns out to be 1.23~MeV, somewhat larger than the only
available experimental information quoting a value of 
$\Gamma_{a_1\to \pi \gamma}=(0.64\pm 0.24)$~MeV. This can 
be substantially improved upon with a reduced
cutoff $\Lambda_{\rho\pi a_1}=1$~GeV, requiring a slightly higher
coupling constant $G_{\rho\pi a_1}=13.27$~GeV$^{-1}$ to fit the
hadronic decay width, but resulting in a photon decay width of
$\Gamma_{a_1\to\pi\gamma}=0.66$~MeV, now in good agreement with
the experimental value. 

The $K_1(1270)$, which was also
included in Ref.~\cite{RCW}, is the appropriate
resonance in $\rho K$ scattering. Since there is no radiative
decay known, we assume (in analogy to the $a_1$ 
cutoff) $\Lambda_{\rho KK_1}=1$~GeV. Taking
$\Gamma_{K_1\to K\rho}=60$~MeV, $m_{K_1}=1270$~MeV yields $G_{\rho
K K_1}=9.42$~GeV$^{-1}$, resulting in $\Gamma_{K_1\to
K\gamma}=0.32$~MeV, which is qualitatively in line with the naive
expectation that in comparison to the $a_1$, the heavier strange
quark essentially acts as a spectator in the photon decay. 

The
$h_1(1170)$ is the isospin-0 pendant to the $a_1(1260)$. No
quantitative empirical information on its decay properties is
available. We therefore make the plausible assumption that the
major part of its width originates from the only observed
$\rho\pi$ decay channel, {\it i.e.}, $\Gamma_{h_1\to\pi\rho}\simeq
300$~MeV. Using again $\Lambda_{\rho\pi h_1}=1$~GeV gives
$G_{\rho\pi h_1}= 11.37$~GeV$^{-1}$ and
$\Gamma_{h_1\to\pi\gamma}=0.6$~MeV, which seems not unreasonable.

The $\omega(782)$ meson differs from the previously discussed
resonances in that it lies significantly below the
free $\pi\rho$ threshold. 
Hadronic models including the effective
four-meson $\omega 3\pi$-vertex usually  attribute substantial
parts of the hadronic decay width of about 7.5~MeV to $\pi\rho$ 
states~\cite{gsw62}. On the other hand, the radiative
decay $\omega\to\pi\gamma$ should, within VDM, entirely proceed
through $\pi\rho$ states. Enforcing the experimentally rather
precisely known value of $\Gamma_{\omega\to\pi\gamma}=0.72$~MeV,
and using again $\Lambda_{\rho\pi\omega}=1$~GeV, yields the
coupling constant $G_{\rho\pi\omega}= 25.8$~GeV$^{-1}$, entailing
$\Gamma_{\omega\to\pi\rho}=3.5$~MeV, which is somewhat on the low
side of the typical values~\cite{gsw62,bali95}. One could
accommodate larger values by choosing softer form factors, but more 
detailed information should be inferred from the dalitz decay spectrum 
$\omega\to\pi^0\mu^+\mu^-$, see, {\it e.g.}, Ref.~\cite{KKW96}.
Let us point out here
that the Dalitz decay $\Gamma_{\omega\to \pi^0e^+e^-}$ is about
one order of magnitude larger than the direct dilepton decay width
$\Gamma_{\omega\to e^+e^-}$~\cite{PDG}. 

The $f_1(1285)$ is
similar to the $\omega(782)$ in the sense that it is also a
sub-threshold state (about 250~MeV below twice the $\rho$ mass),
accompanied by a rather small total decay width of $\sim$~25~MeV; this
supports  a tempting interpretation as a state with a large $\rho\rho$
component, the decay into it being suppressed by the lack of phase
space. We will have more to say on this later. For the time being, if 
we attribute the entire decay width into 4
pions of about 7.5~MeV to the $\rho\rho$ channel, the radiative
decay is overestimated by at least a factor of 2. Smaller form
factor cutoffs are not really efficient in suppressing the
radiative decay width, since the latter still involves an integral
over one $\rho$ spectral function where dominant contributions
arise from masses $M_2\simeq m_\rho$, where the decay momentum is
rather small. A reasonable compromise for our purposes appears to
fix the radiative decay width approximately at its experimental
value of $\Gamma_{f_1\to\rho^0\gamma}\simeq 1.65$~MeV, which,
choosing a form factor cutoff $\Lambda_{\rho\rho f_1}=0.8$~GeV,
results in a hadronic decay width $\Gamma_{f_1\to\rho\rho}=3$~MeV.

The total width of the $\pi(1300)$ is not very well known,  
quoted as 200-600~MeV by the PDG~\cite{PDG}. Lacking more precise
information, we attribute 300~MeV to the $\pi\rho$ channel, which
is one of the two observed decay modes. The specific form of our 
interaction Lagrangian, Eq.~(\ref{LrhoPP}), together with 
the VDM assumption, does not allow any radiative decay. The latter
has not been observed so far. 

The coupling constants and cutoff parameters as well as the 
resulting branching ratios are summarized in Table~\ref{tabfit}.

\section{In-Medium $\rho$ Propagator}
\label{sec_rhoprop}
Having fixed the parameters of the interaction vertices, we are now
set to calculate the in-medium $\rho$ self-energy, $\Sigma_\rho$,
and corresponding dilepton production rates. The self-energy is related
in a standard way to the forward two-body $\rho$ scattering amplitude
off the surrounding thermal mesons. Within the imaginary time
(Matsubara) formalism one obtains:
\begin{equation}
\Sigma_{\rho h}^{\mu\nu}(q_0,\vec q;T)=\int \frac{d^3p}{(2\pi)^3}
\frac{1}{2\omega_h(p)} [f^h(\omega_h(p))-
f^{\rho h}(\omega_h(p)+q_0)] \ M_{\rho h}^{\mu\nu}(p,q) \ ,
\label{sigmamunu}
\end{equation}
where the isospin averaged $\rho$ scattering
amplitude $M_{\rho h}$ is integrated over the thermal Bose distribution
$f^h(\omega_h(p))=[\exp(\omega_h(p))/T-1]^{-1}$  of the corresponding
hadron species $h$ with $\omega_h(p)=\sqrt{m_h^2+\vec p^2}$.
The invariant amplitudes are evaluated in terms of the
$s$-channel resonance contributions from the previous section.
Collisions of the $\rho$ with the pseudoscalar mesons $P=\pi, K$ lead to
\begin{eqnarray}
M_{\rho PA}(p,q) &=&
IF \ G_{\rho PA}^2 \ F_{\rho PA}(q_{cm})^2
(\varepsilon^{\kappa} \ p \cdot q- q^\kappa \ p\cdot\varepsilon)
 \ D_{A,\kappa\lambda}(s) \ (\varepsilon^{*\lambda} \ p \cdot q -
q^\lambda \ p\cdot\varepsilon^*) \ .
\nonumber\\
\end{eqnarray}
Then, one obtains for the axial-vector resonances $A=h_1,a_1,K_1$: 
\begin{eqnarray}
\quad M_{\rho PA}^{\mu\nu}(p,q) &=& IF \ G_{\rho PA}^2 \
F_{\rho PA}(q_{cm})^2 \ D_A(s) \ v_A^{\mu\nu}(p,q) \
\nonumber\\
v_A^{\mu\nu}(p,q) &=&
-g^{\mu\nu} (p\cdot q)^2+q^\mu q^\nu \frac{(p\cdot q)^2}{s}
+(q^\mu p^\nu +q^\nu p^\mu) p\cdot q (1-\frac{q^2}{s})
+ p^\mu p^\nu q^2 (1-\frac{q^2}{s}) \ , 
\nonumber\\
\label{MrhoPA}
\end{eqnarray}
for the vector resonance $V=\omega$: 
\begin{eqnarray}
 M_{\rho PV}(p,q) & = &
IF \  G_{\rho PV}^2 \ F_{\rho PV}(q_{cm})^2
(\epsilon^{\alpha\kappa\beta\mu} \ \varepsilon_\mu \ k_\alpha \ q_\beta)
 \ D_{V,\kappa\lambda}(s) \
(\epsilon^{\gamma\lambda\delta\nu} \ \varepsilon_\nu^* \ k_\gamma \ q_\delta)
\nonumber \ ,\\
M_{\rho PV}^{\mu\nu}(p,q) & = & IF \ G_{\rho PV}^2 \
F_{\rho PV}(q_{cm})^2 \ D_V(s) \ v_V^{\mu\nu}(p,q) \ ,
\nonumber\\
v_V^{\mu\nu}(p,q) & = & -g^{\mu\nu} ((p\cdot q)^2-p^2 q^2) -q^\mu q^\nu p^2
+(q^\mu p^\nu + q^\nu p^\mu) p\cdot q - p^\mu p^\nu q^2 \ , 
\label{MrhoPV}
\end{eqnarray}
and for the  pseudoscalar resonance $\pi'(1300)$:  
\begin{eqnarray}
M_{\rho PP'}(p,q) &=&
IF  G_{\rho PP'}^2 F_{\rho PP'}(q_{cm})^2
(\varepsilon\cdot p \ k\cdot q - \varepsilon\cdot k \ p\cdot q) 
 \ D_{P'}(s) \
(\varepsilon^*\cdot p \ k\cdot q - \varepsilon^*\cdot k \ p\cdot q), 
\nonumber\\
M_{\rho PP'}^{\mu\nu}(p,q) &=& IF \ G_{\rho PP'}^2 \
F_{\rho PP'}(q_{cm})^2 \ D_{P'}(s) \ v_{P'}^{\mu\nu}(p,q) \ ,
\nonumber\\
v_{P'}^{\mu\nu}(p,q) &=&  q^\mu q^\nu (p\cdot q)^2
-(q^\mu p^\nu + q^\nu p^\mu) \ q^2 \ p\cdot q + p^\mu p^\nu q^4 \ . 
\label{MrhoPP}
\end{eqnarray}
The amplitude tensors $M^{\mu\nu}$
are obtained by removing the $\rho$ meson polarization vectors
$\varepsilon_\mu, \varepsilon^*_\nu$ from the invariant matrix elements and
contracting the remaining indices.
The intermediate (axial-) vector propagators at four-momentum $k\equiv(p+q)$
have been taken as ($R=A,V$)
\begin{eqnarray}
D_{R,\kappa\lambda}(k) & = &
\frac{(-g_{\kappa\lambda}+k_\kappa k_\lambda / s)}
{s-m_{R}^2+im_{R}\Gamma_{R}^{tot}(s)} \
\nonumber\\
 & \equiv & (-g_{\kappa\lambda}+k_\kappa k_\lambda / s) \ D_{R}(s) \ .
\end{eqnarray}

Using the same form of propagator for the $f_1$-meson in the
$\rho\rho$ scattering process raises a problem: the corresponding (Born-)  
amplitude for $\rho\rho \to f_1 \to \rho\rho$ vanishes identically, which
is, in fact, in line with analyses based on chiral 
Lagrangians~\cite{bali95,gomm}. However, 
the latter then suggests, within VDM, the absence of the direct 
$f_1\to \rho^0 \gamma$ decay, which empirically is quite large. 
Since here we are interested 
in phenomenological estimates for dilepton production, we decided to 
circumvent the vanishing $\rho\rho f_1$ coupling  by making use of the gauge 
freedom for massive vector particles provided by the additional 
$-\frac{\lambda}{2}(k_\beta f_1^\beta)$-term (St\"uckelberg term~\cite{ItZu})
in the interaction Lagrangian, Eq.~(\ref{LrhoVA}). With $\lambda=1$ 
the $f_1$ propagator takes the form 
\begin{equation}
D_{f_1}^{\kappa\lambda}(s)=-g^{\kappa\lambda} D_{f_1}(s) \ , 
\end{equation} 
which has been used in the actual calculations. We note that when further
applying the naive VDM for the two-photon decay of the $f_1$, we would 
obtain a nonzero branching ratio, violating Yang's theorem~\cite{yang}.
This is, of course, an artefact that might be related to our gauge 
choice (in fact, Brihaye {\it et al.}~\cite{Briha} pointed out 
that VDM breaks down at the two-photon level). 
On the other hand, we have found that the $f_1$ contributes 
negligibly to the quantities being calculated in this paper, with the 
parameters delineated as described above. Therefore, the role of the  
$f_1$, which at first seemed promising because of its large radiative
decay width, turned out to be  unimportant from a pragmatic point of 
view, so that we
do not attempt further improvements of its interaction vertex.   

In the medium, the specification of a thermal rest frame breaks Lorentz
invariance. As a consequence, the in-medium $\rho$ self-energy tensor
is characterized by two independent scalar functions, each depending
separately on energy and three-momentum (in the vacuum one scalar
function depending on invariant mass only is sufficient). This is
conveniently described in terms of longitudinal and transverse modes
of the $\rho$ propagator~\cite{GK91}:
\begin{equation}
D_\rho^{\mu\nu}(q_0,\vec q) = \frac{P_L^{\mu\nu}}{M^2-(m_\rho^0)^2
-\Sigma_\rho^L(q_0,\vec q)}+\frac{P_T^{\mu\nu}}{M^2-(m_\rho^0)^2
-\Sigma_\rho^T(q_0,\vec q)} + \frac{q^\mu q^\nu}{(m_\rho^0)^2 M^2} 
\label{drhomunu}
\end{equation}
with the standard projection operators
\begin{eqnarray}
P_L^{\mu\nu} & = & \frac{q^\mu q^\nu}{M^2}-g^{\mu\nu}-P_T^{\mu\nu}
\nonumber\\
P_T^{\mu\nu} & = &  \left\{ \begin{array}{l}
 \quad~~ 0 \qquad , \mu=0 \ {\rm or} \ \nu=0
 \\
\delta^{ij}-\frac{q^iq^j}{{\vec q}^2} \ , \ \mu,\nu \in \lbrace 1,2,3
\rbrace
\end{array}   \right . 
\label{PLT}
\end{eqnarray}
(the spacelike components of $\mu$ and $\nu$ are denoted by
$i$ and $j$, respectively).
The longitudinal and transverse self-energies are defined by the
corresponding decomposition of the polarization tensor:
\begin{equation}
\Sigma_\rho^{\mu\nu}(q_0,\vec q)= \Sigma_\rho^L(q_0,\vec q) \
P_L^{\mu\nu} + \Sigma_\rho^T(q_0,\vec q) \ P_T^{\mu\nu} \ ;
\end{equation}
they are calculated from Eq.~(\ref{sigmamunu}) as
\begin{eqnarray}
\Sigma_{\rho PR}^{L,T}(q_0,q) &=& G_{\rho PR}^2 \ IF \ 
\int\frac{\vec p^2 d|\vec p| dx}{(2\pi)^2 2\omega_P(p)}
 \ [f^P(\omega_P(p))-f^{\rho P}(\omega_P(p)+q_0)] \
 F_{\rho PR}(q_{cm})^2 
\nonumber\\ 
 & & \qquad \qquad \times \ D_{R}(s) \  v_{R}^{L,T}(p,q)
\end{eqnarray}
for $R=A,V,P'$ with $x=\cos\theta$, $\theta=\angle(\vec p, \vec q)$. The  
projected vertex functions are given by 
\begin{eqnarray}
v_A^L(p,q) & = & (P_L)_{\mu\nu} \ v_A^{\mu\nu}(p,q)
\nonumber\\
 & = & \frac{q^2}{4s}(s-q^2-m_\pi^2)^2
+ {\omega_P(p)}^2 q^2 (1-\frac{q^2}{s})
-\vec p^2 q^2 x^2 (1-\frac{q^2}{s})
\nonumber\\
v_A^T(p,q) & = & \frac{1}{2} (P_T)_{\mu\nu} \ v_A^{\mu\nu}(p,q)
\nonumber\\
 & = & \frac{1}{2} \ [\frac{1}{2}(s-q^2-m_\pi^2)^2
-\vec p^2 q^2 (1-x^2) (1-\frac{q^2}{s})]
\nonumber\\
v_V^L(p,q) & = & (P_L)_{\mu\nu} \ v_V^{\mu\nu}(p,q)
\nonumber\\
 & = & q^2 \vec p^2 (1-x^2)
\nonumber\\
v_V^T(p,q) & = & \frac{1}{2} (P_T)_{\mu\nu} \ v_V^{\mu\nu}(p,q)
\nonumber\\
 & = &  \frac{1}{2} \ [\frac{1}{2}(s-q^2-m_\pi^2)^2
-2m_P^2 q^2-q^2 \vec p^2 (1-x^2)] \ 
\end{eqnarray}
and similar expressions for the pseudoscalar resonance $P'$.  
For the $f_1(1285)$ an additional integration over the mass distribution
of the $\rho$ from the heat bath has to be performed:
\begin{eqnarray}
\Sigma_{\rho\rho f_1}^{L,T}(q_0,q) &=&
G_{\rho\rho f_1}^2 \ \int \frac{M_2 dM_2}{\pi} A_\rho(m_2)
\int\frac{\vec p^2 d|\vec p| dx}{(2\pi)^2 2p_0}
 \ [f^\rho(p_0)-f^{\rho\rho}(p_0+q_0)] \
 F_{\rho\rho f_1}(q_{cm})^2 \ 
\nonumber\\ 
 & & \qquad\qquad \times \ D_{f_1}(s) \  v_{f_1}^{L,T}(p,q) \ ,
\end{eqnarray}
where $M_2^2=p_0^2-\vec p^2$. 
\begin{figure}[htb]
\hspace{0.8in}
\epsfig{figure=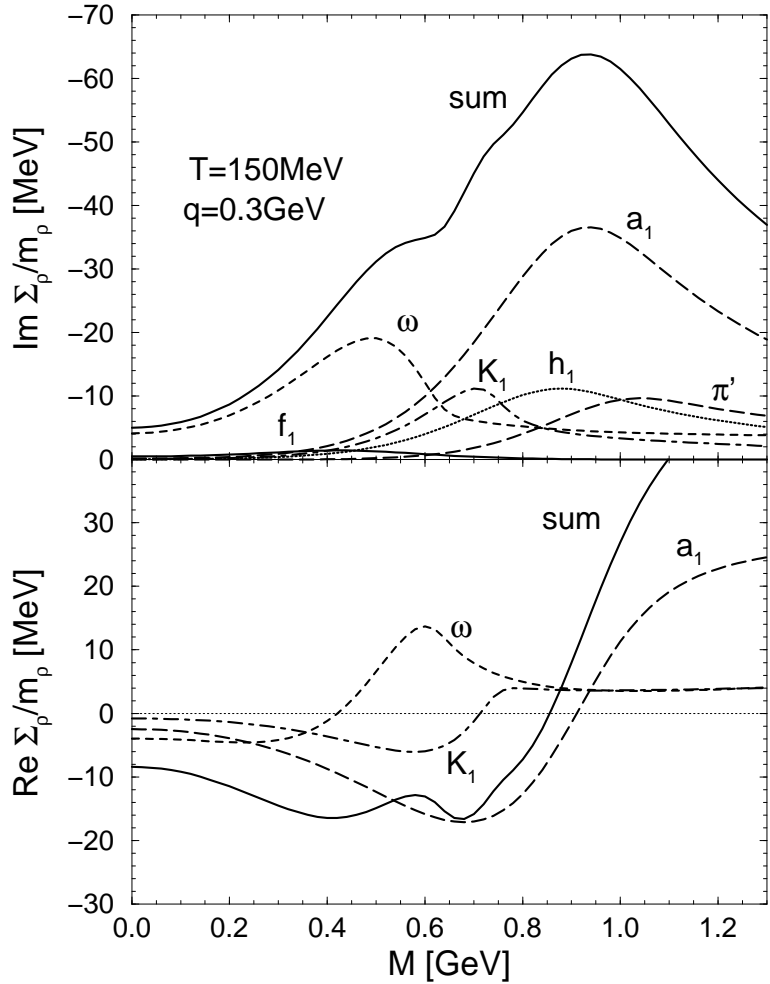,height=5.5in,width=4in}
\caption{The real and imaginary parts of the polarization-averaged
$\rho$ self-energy (lower and upper panel, respectively). The
different channels are labelled explicitly and explained in the text.
Note that the $\pi\pi$ channel is absent for the sake of clarity.}
\label{fig_selfe}
\end{figure}
In Fig.~\ref{fig_selfe} the real and imaginary parts of the individual
spin-averaged self-energy contributions, 
\begin{equation}
\Sigma_{\rho hR}(M,q)=\frac{1}{3} 
\left[ \Sigma_{\rho hR}^L(M,q)+2\Sigma_{\rho hR}^T(M,q) \right] \ , 
\end{equation} 
($h=\pi, K, \rho$) are shown at fixed three-momentum modulus 
$|\vec q|=0.3$~GeV 
in the lower and upper panel, respectively. Around and 
above the free mass $m_\rho$, the strongest absorption is caused by 
$a_1(1260)$ resonance formation, which is about as large as the sum of 
all other channels, shared to roughly equal amounts between 
$K_1(1270)$, $h_1(1170)$ and $\pi'(1300)$. The $K_1(1270)$ curve 
acquires its maximum at lower $M$ than the pion-resonances due to 
the higher thermal energies of the kaons (including their rest mass). 
In the low-mass region $M\le 0.6$~GeV, the 
dominant contribution is due to the $\omega$-meson, which, however,  
barely leaves any trace in the resonance region. 
It is also seen that the effect of the $f_1$ is very small.
In the real part of the 
total self-energy we observe appreciable cancellations,  until eventually all
contributions turn repulsive (the latter feature would of course be
modified when accounting for further higher resonances). 
Such cancellations are typical for
many-body type calculations as performed here. They are the reason that one
usually encounters only moderate modifications
of the in-medium pole mass. On the other hand, the imaginary parts of
$\Sigma_\rho$ strictly add up, generating significant broadening.  We
refrain here from plotting the two-pion loop contribution to the $\rho$
self-energy. It will start exceeding the in-medium corrections past
$M\simeq 0.450$~GeV (the free $\rho$ width at $M=m_\rho$
amounts to about 150~MeV).

\begin{figure}
\vspace{-0.2in}
\hspace{0.3in}
\epsfig{figure=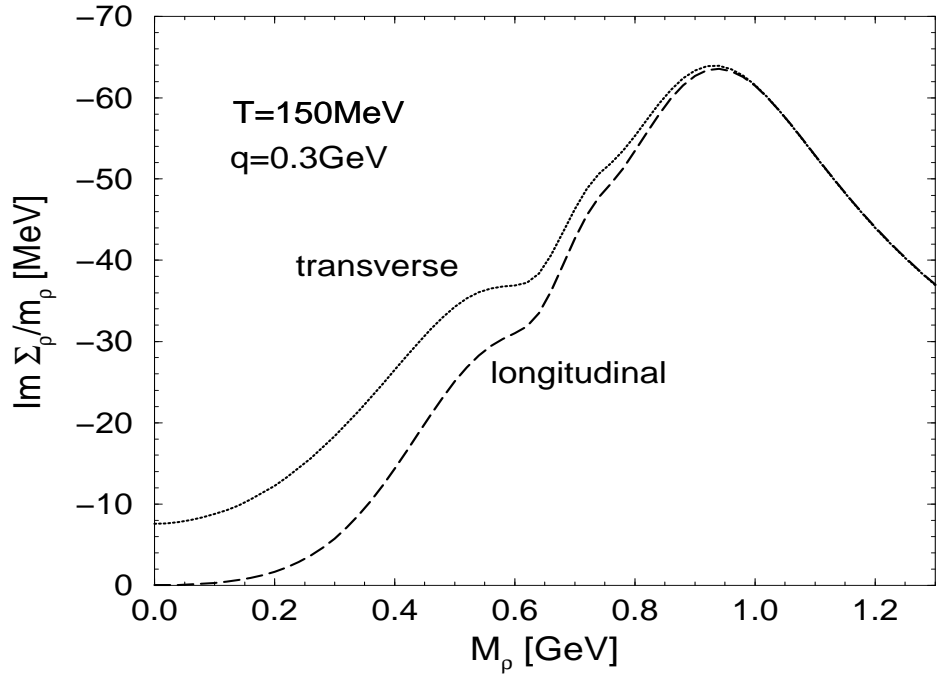,height=5.5in,width=4in,angle=-90}
\caption{The longitudinal and transverse polarization
contributions to the imaginary part of the self-energy as arising
from the sum of meson resonance contributions.}
\label{polariz}
\end{figure}

As discussed earlier, the existence of a preferred thermal 
reference frame will break Lorentz invariance. An advantage of a
theoretical approach like the one at hand is that the transverse and
longitudinal parts of the rho self-energy can be separately
resolved. This is shown on Fig.~\ref{polariz}. Even though at the present time
one does not have a practical observable that is convincingly sensitive to
the 
polarization, one should keep this difference in mind for future
applications~\cite{GK91}. From the figure, we see that at finite 
three-momentum the
polarizations differ the most at low invariant masses and become
undistinguishable at high masses.

In addition to the direct $\rho-h$ interactions 
we account for the simplest medium effect in the pion
cloud of the $\rho$ by including the Bose enhancement in the two-pion
bubble. In the Matsubara formalism this amounts two
replacing the two-pion propagator, Eq.~(\ref{Gpipi}), by~\cite{HK87}
\begin{equation}
G_{\pi\pi}(M,p;T)=\frac{1}{\omega_\pi(p)} \
\frac{\left[1+2f^\pi(\omega_\pi(p);T)\right]}{M^2-(2 \omega_\pi(p))^2+i\eta}
\label{GpipiT}
\end{equation}
At a temperature of $T=150$~MeV, this generates an additional 
broadening of the $\rho$ self-energy starting from the two-pion
threshold  reaching an appreciable  maximum of $\sim$~20~MeV
at $M\simeq 0.6$~GeV (on the scale of Fig.~1, upper panel) and  
gradually decreasing beyond. 

\begin{figure}
\vspace{-0.5in}
\hspace{0.3in}
\epsfig{figure=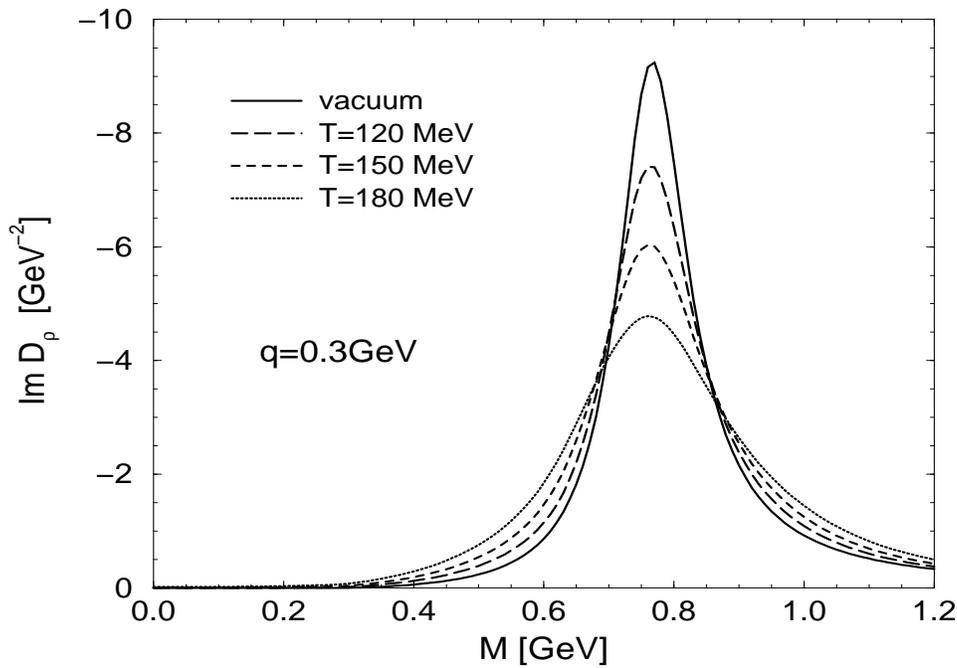,height=5.5in,width=4in,angle=-90}
\vspace{0.2in}
\caption{Imaginary part of the $\rho$-propagator (spectral function)
 in the vacuum (dotted curve), and in a thermal
gas including the full in-medium self-energies, Eq.~(\ref{selfep}), for fixed
three-momentum $q=0.3$~GeV  at temperatures $T=120$~MeV (long-dashed curve),
$T=150$~MeV (dashed curve) and $T=180$~MeV (dotted curve).}
\label{fig_spectral}
\end{figure}

To end this section, we plot in Fig.~\ref{fig_spectral} the full spin-averaged 
imaginary part of the $\rho$ propagator (spectral function), 
\begin{equation}
{\rm Im} D_\rho(M,q;T) =\frac{1}{3} \left[{\rm Im} D_\rho^L(M,q;T)+
2{\rm Im} D_\rho^T(M,q;T) \right]
\label{imdrho}
\end{equation}
in a thermal meson gas of temperatures $T=120$, 150 and 180~MeV as 
appropriate for the hadronic phase in ultrarelativistic heavy-ion 
collisions. More explicitly, one has 
\begin{equation}
{\rm Im}D_\rho^{L,T}(M,q;T)=
\frac{{\rm Im} \Sigma_\rho^{L,T}(M,q;T)}{|M^2-(m_\rho^{0})^2
-\Sigma_\rho^{L,T}(M,q;T)|^2}  \ 
\label{imdrhoLT}
\end{equation}
with the longitudinal and transverse self-energy parts 
\begin{eqnarray}
\Sigma_\rho^L & = & \Sigma_{\rho\pi\pi} + \sum\limits_{\alpha}
\Sigma_{\rho\alpha}^L
\nonumber\\
\Sigma_\rho^T & = & \Sigma_{\rho\pi\pi} +\sum\limits_{\alpha}
\Sigma_{\rho\alpha}^T \ ,
\label{selfep}
\end{eqnarray}
where the summation is over the mesonic excitation channels  
$\alpha$=$\pi\omega$, $\pi h_1$, $\pi a_1$, $\pi\pi'$, $KK_1$, 
$\bar K \bar K_1$, $\rho f_1$, as discussed, 
and $\Sigma_{\rho\pi\pi}$ now contains the 
Bose-Einstein factors through Eq.~(\ref{GpipiT}). 
We find that the thermal $\rho$ spectral function undergoes
a broadening (defined as the full width at half maximum) of about 
80~MeV at $T=150$~MeV (with little three-momentum dependence, see also, 
{\it e.g.}, Ref.~\cite{RCW}),
 which almost doubles to $\sim 155$~MeV at 
$T=180$~MeV. Those values are a factor of 2 larger than the collisional 
broadening found in Ref.~\cite{ha95} based on on-shell scattering 
amplitudes. In Ref.~\cite{EIK98} the $\rho$ meson self-energy has also
been evaluated for on-shell $\rho$ mesons using the 
$T_{\rho h}$-$\varrho_h$ approximation ({\it i.e.}, the self-energy being  
proportional to the $\rho$-$h$ 
scattering amplitude and the matter particle density $\varrho_h$). 
For a pion gas of density $n_\pi=1.5$~fm$^{-3}$ a broadening of 400~MeV
has been quoted, which, when rescaling to a density of 0.12~fm$^{-3}$
(corresponding to thermal equilibrium at $T=150$~MeV) gives $\sim$~30~MeV, 
again about a factor 2 smaller than our results; this is not
surprising as the meson resonances included in ref.~\cite{EIK98} were
the $a_1(1260)$, $\pi'(1300)$, $a_2(1320)$ and $\omega(1420)$, the latter
three contributing rather little at the free $\rho$ mass $M=m_\rho$. 
On the other hand, the recent kinetic theory treatment performed
in Ref.~\cite{gao98} does agree with our findings. However, 
we would like to stress again that our
approach consistently accounts for the empirical radiative decays 
at the same time, which is crucial for reliable predictions of 
low-mass dilepton production to be addressed in the next section. 
The shift of the pole mass, defined by the zero crossing in the real part
of the propagator, turns out to be negligible, moving from 
$M=773$~MeV in vacuum to $M=776$~MeV at $T=150$~MeV. 

\section{Dilepton Production}
\label{sec_dilep}
The differential dilepton production rate per unit four-volume and
four-momentum in hot matter can be decomposed as~\cite{GK91} 
\begin{equation}
\frac{dN_{l^+l^-}}{d^4xd^4q}=L_{\mu\nu}(q) H^{\mu\nu}(q) \ . 
\end{equation}
For definiteness, we will focus on $e^+e^-$ pairs in the following.
Then the electron/positron  rest masses can be neglected as compared to
their three-momenta, $m_{e^\pm}\ll |\vec p_\pm|$, and,
to lowest order in the electromagnetic coupling $\alpha$, 
the lepton tensor takes the form
\begin{equation}
L_{\mu\nu}(q)=-\frac{\alpha^2}{3\pi^2 M^2} \left( g_{\mu\nu} -
\frac{q_\mu q_\nu}{M^2} \right) \
\end{equation}
with the total pair four-momentum $q=p_++p_-$. Here, we are interested
in dilepton production from $\rho$ decays (or, equivalaently, $\pi\pi$
annihilation).
Within the VDM, which we
have already employed in the evaluation of radiative decay widths
in sect.~\ref{sec_lagr}, the corresponding hadronic tensor is directly related to
the imaginary part of the retarded $\rho$ propagator in hot and
dense matter:
\begin{equation}
H^{\mu\nu}(q_0, \vec q;\mu_B,T)=-f^\rho(q_0;T) \ \frac{(m_\rho^0)^4}
{\pi g_{\rho\pi\pi}^2} \ {\rm Im}D_\rho^{\mu\nu}(q_0,\vec q;\mu_B,T) \ . 
\end{equation}
Using the decomposition Eq.~(\ref{drhomunu}), the dilepton rate can then
be written as
\begin{equation}
 {dN \over d^4xd^4q} =
-\frac{\alpha^2 (m_\rho^{0})^4}{\pi^3 g_{\rho\pi\pi}^2} \
\frac{f^\rho(q_0;T)}{M^2} \ \frac{1}{3} \
\left[ {\rm Im}D_\rho^L(q_0,q;T)+2 \ {\rm Im}D_\rho^T(q_0,q;T)\right]
\label{rate}
\end{equation}
with the longitudinal and transverse spectral functions given
by Eq.~(\ref{imdrhoLT}).  

\begin{figure}
\vspace{-0.5in}
\hspace{0.3in}
\epsfig{figure=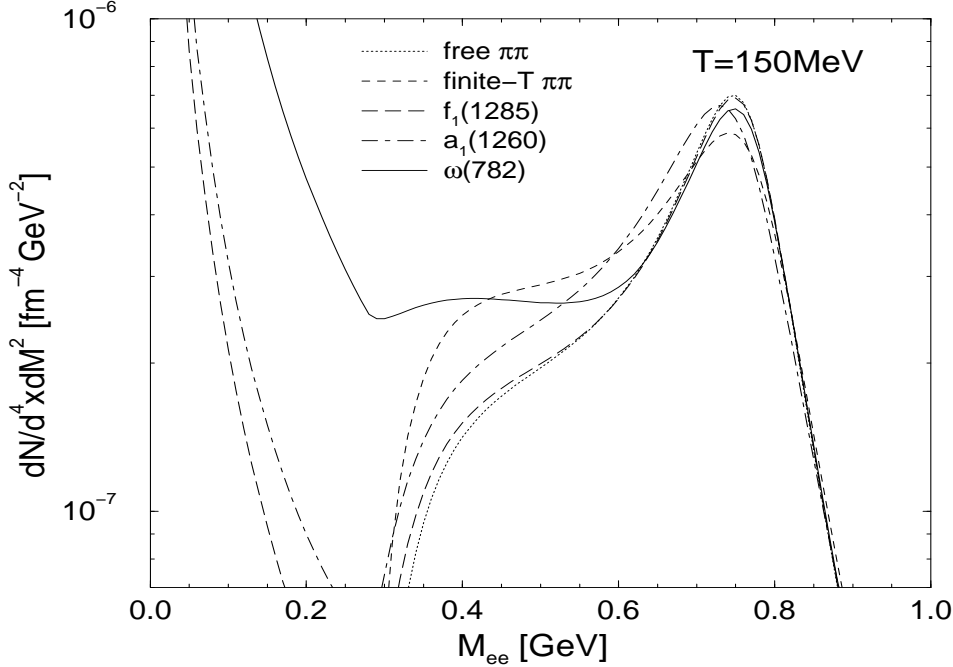,height=5.5in,width=4in,angle=-90}
\vspace{0.2in}
\caption{Three-momentum integrated dilepton production rates
at temperature of $T=150$~MeV including various
individual medium effects in the $\rho$ propagator
(see Eqs.~(\protect\ref{rate}),
(\protect\ref{imdrhoLT}), (\protect\ref{selfep}));
dotted curve: free $\pi\pi$ annihilation
({\it i.e.}, using $\Sigma_\rho^{L,T}=\Sigma_{\rho\pi\pi}^0$);
short-dashed curve: $\pi\pi$ annihilation including finite temperature
effects through Bose enhancement factors,  Eq.~(\protect\ref{GpipiT})
(using $\Sigma_\rho^{L,T}=\Sigma_{\rho\pi\pi}$);
long-dashed curve: free $\pi\pi$ including $\rho\rho\to f_1$ resonance
formation (using
$\Sigma_\rho^{L,T}=\Sigma_{\rho\pi\pi}^0+\Sigma_{\rho\rho f_1}^{L,T}$);
dashed-dotted curve: free $\pi\pi$ including $\rho\pi\to a_1$ resonance
formation (using
$\Sigma_\rho^{L,T}=\Sigma_{\rho\pi\pi}^0+\Sigma_{\rho\pi a_1}^{L,T}$);
solid curve: free $\pi\pi$ including $\rho\pi\to \omega$ resonance
formation (using
$\Sigma_\rho^{L,T}=\Sigma_{\rho\pi\pi}^0+\Sigma_{\rho\pi\omega}$).}
\label{fig_rates}
\end{figure}

In Fig.~\ref{fig_rates} we display the individual medium effects in the 
three-momentum integrated rates
\begin{equation}
\frac{dN}{d^4x dM^2}(M;T)
=\int\frac{d^3q}{2q_0} \frac{dN}{d^4xd^4q}(q_0,\vec q) \ . 
\label{rate2} 
\end{equation} 
for electron-positron production at a fixed temperature of $T$=150~MeV. 
The two mechanisms that  dominate the net 
\noindent low-mass dilepton rate are  associated with 
$\rho\pi \to \omega$ formation as well as the 
Bose-Einstein (BE) enhancement of the $\rho\to \pi\pi$ decay width.  
{\it E.g.}, at $M$=0.4~GeV, both the $\omega$ (solid line) and the 
finite-$T$ (BE) corrections in the pion cloud enhance the free
$\pi\pi$ annihilation rate by (80-90)\% each. Much smaller effects
are due to the $a_1(1260)$ (30\% enhancement, dashed-dotted line)
as well as the $K_1$ and $h_1$ (20\% and 15\% enhancement, respectively, 
not shown in Fig.~4), the $f_1$ being practically negligible.  

\begin{figure}
\vspace{-0.5in}
\hspace{0.3in}
\epsfig{figure=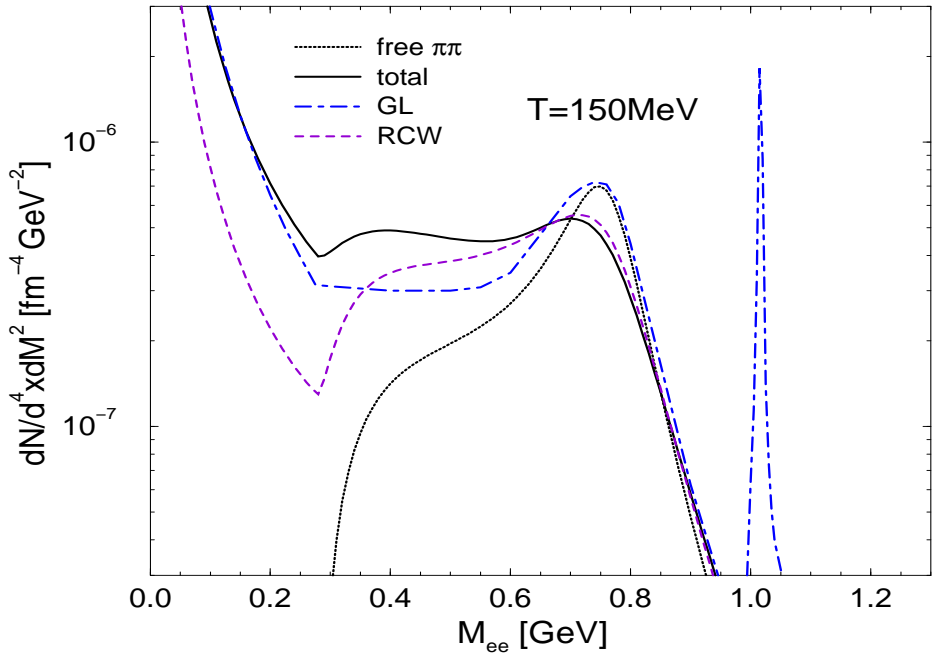,height=5.5in,width=4in,angle=-90}
\vspace{0.2in}
\caption{Comparison of our total, three-momentum integrated thermal dilepton
production rate (solid curve)
with that obtained from a sum of meson decays and reactions according to
Ref.~\protect\cite{gali} (dashed-dotted curve).
The dotted curve represents
the rate obtained from free $\pi\pi$ annihilation, whereas the dashed curve
corresponds to the finite temperature part of the many-boby calculation
of Ref.~\protect\cite{RCW}.  All curves are for a fixed temperature 
$T=150$~MeV.}
\label{fig_gl}
\end{figure}

In Fig.~\ref{fig_gl} we compare the total rate
calculated from an incoherent sum of meson reactions 
and decays~\cite{gali} (dashed-dotted line) 
with our full result (solid line). Focusing again at $M=0.4$~GeV, the 
latter is increased over the free $\pi\pi$ result (=GL, dotted line) 
by a factor of 3.5, compared to $\sim$~2 in the Gale-Lichard
calculation. The major difference  arises from the inclusion of the 
BE-enhancement (and, to a lesser extent, the $a_1/h_1$ resonances) 
in the present treatment. Also shown in Fig.~\ref{fig_gl} are the results
based on the finite temperature part of the calculation by Rapp, Chanfray
and Wambach (=RCW, short-dashed curve)~\cite{RCW}, which included 
$a_1$, $K_1$ resonances and the 
Bose-Einstein enhancement in $\Sigma_{\rho\pi\pi}$ (note, however, 
that the hadronic form factors used in Ref.~\cite{RCW} were 
substantially harder, with cutoffs 
$\Lambda_{\pi\rho a_1}=\Lambda_{K\rho K_1}=2$~GeV; this 
entails an overestimation of the {\em radiative} 
$a_1\to\gamma\pi$ decay width by a factor of 2, as elaborated in
sect.~\ref{sec_param}). 
Consequently, in the vicinity of the free $\rho$ mass, where
the major broadening effect is due to $a_1$ formation and BE-enhancement, 
the RCW results differ very little from the present ones. Below, $\omega$ 
formation in $\rho\pi$ scattering is responsible for a substantial increase
of the emission rate (full curve compared to short-dashed curve). 

Another point is finally noteworthy of attention: if
one looks at the total spectrum in the vicinity of the $rho$-peak, 
one sees that the net
signal in our many-body spectral function approach is  
reduced by about 40\% as compared to free $\pi\pi$ annihilation.  
As discussed in the previous section, the shift of the real part
is very small (a few MeV; 
the {\em apparent} larger shift in the dilepton spectrum
is caused by the overall Bose factor in the rate expression, 
Eq.~(\ref{rate}), 
which strongly increases towards smaller invariant masses), 
but the broadening of $\sim$~80~MeV is appreciable   
on the scale of the free $\rho$ width of $\sim$~150~MeV. 
The resulting peak smearing is a distinct  
feature of our many-body formalism, {\it i.e.}, the resummation 
encoded in  the  denominator in Eq.~(\ref{imdrhoLT}), which is neither  
present in the
rate calculations of Ref.~\cite{gali} nor in low-density expansions
as performed, {\it e.g.}, in Ref.~\cite{SYZ1}.  
This feature will be tested against experiment soon, as a result
of improving the mass resolution in the CERES experiment at 
CERN-SpS energies, as well as at RHIC energies in the PHENIX detector,
where the design value for $\Delta M/M$ is at the 1\% level.   

\section{Summary and Conclusions}
\label{sec_concl}
Based on a finite temperature (Matsubara) formalism  we have 
highlighted different interaction channels that could lead to
modifications of the rho properties in a thermal gas of mesons with  
zero chemical potentials. 
Using phenomenological interaction vertices compatible with 
gauge invariance and chiral symmetry, and including hadronic 
form factors to simulate finite size effects, 
the free parameters could be tuned to reproduce the empirical 
hadronic {\em and} radiative decay branchings rather well. 
The resulting in-medium $\rho$ self-energy induces  a moderate  
broadening of the $\rho$ spectral function, somewhat higher than what 
has been found before in Refs.~\cite{ha95,RCW,EIK98}, but consistent with 
a recent kinetic theory analysis~\cite{gao98}. On the other hand, 
 cancellations in the real
part inhibited significant changes of the in-medium $\rho$ pole-mass.  
Corresponding dilepton spectra exhibit a 40\%  
depletion of the $\rho$-peak together with an appreciable enhancement of
a factor of $\sim$~3.5 below, largely driven by sub-threshold 
$\pi\rho\to \omega$
formation and a Bose enhancement in the $\rho\to\pi\pi$ width. 
Our results have now to be combined with those obtained
with the baryons present, and a time-evolution approach will be coupled
with our rates to produce yields that can be compared with 
experiment. 

Finally, for the sake of comparison, we have overlayed our results with 
the those obtained from an (incoherent)  sum of
meson reactions and decays~\cite{gali}. Some differences
emerge: the earlier calculations in Ref.~\cite{gali} do not include 
the contribution that can be associated with the radiative decay 
of the $a_1(1260)$ and $h_1(1170)$;  also,  the VDM form
factor used there was temperature-independent. The methods employed in this
work are tantamount to the generation of a temperature-dependent form
factor. Besides that, it seems that the self-consistent many-body treatment
does not induce large deviations as compared with 
the results of (incoherent) rate calculations as far as the low-mass
enhancement is concerned.  There are, of course, significant differences
around the free $\rho$ mass, in that the many-body calculations lead to a 
reduction of the $\rho$-peak. In fact, these features can be understood
in a rather transparent way as follows. Schematically, the $\rho$ 
spectral function (which directly governs the dilepton rate) can be
written in terms of the self-energy as 
\begin{equation}
{\rm Im} D_\rho= 
\frac{{\rm Im} \Sigma_\rho}{|M^2-m_\rho^2|^2+|{\rm Im}\Sigma_\rho|^2} \ ,  
\label{schematic} 
\end{equation}
where we have absorbed the real part of the self-energy in the (physical) 
$\rho$ mass $m_\rho$. In the low-mass region, where 
$m_\rho\gg M$ and $m_\rho\gg|{\rm Im} \Sigma_\rho|$, the denominator 
is dominated by $m_\rho$ so that
\begin{equation}
{\rm Im} D_\rho(M \ll m_\rho) 
\propto \frac{{\rm Im} \Sigma_\rho}{m_\rho^4}  \ .  
\end{equation}
Since ${\rm Im} \Sigma_\rho$  corresponds to a summation of scattering
amplitudes times (pion-) density, one immediately recognizes the close
resemblance to kinetic theory or low-density expansion approaches. On 
the other hand, in the vicinity of the $\rho$-peak, where 
$M\simeq m_\rho$, the denominator
in Eq.~(\ref{schematic}) is dominated by ${\rm Im}\Sigma_\rho$ so that
\begin{equation}
{\rm Im} D_\rho(M\simeq m_\rho)\propto \frac{1}{{\rm Im} \Sigma_\rho} \ , 
\end{equation}
demonstrating that the consequence of an increase in density
is a suppression of the maximum, which cannot be straightforwardly casted
in a low-density expansion.  
  
Overall, however, the finite temperature effects found here are still to be 
regarded as rather moderate. 
On the contrary,  the nuclear environment has a much stronger 
impact on the in-medium rho-modifications at comparable
densities~\cite{RCW,KKW97,PPLLM,EIK98,gao98}: {\it e.g.}, at a temperature
$T=160$~MeV, where the thermal pion density, $n_\pi=0.16$~fm$^{-3}$, equals 
normal nuclear density, the $\rho$ spectral function broadens by about
$\Gamma_\rho^{med}(T=160{\rm MeV})\simeq 100~$MeV, which is 
considerably less than in 
 nuclear matter ($\Gamma_\rho^{med}(\rho_N=\rho_0)\ge 300$~MeV,
as extracted, {\it e.g.},  from Refs.~\cite{RUBW,Morio98}). 
From a theoretical point of view this state of affairs may appear puzzling, 
since towards the chiral restoration transition, 
which is to be expected around the temperatures considered here, 
a substantial reshaping of the vector and axial-vector spectral distribution
must occur: for $T\ge T_c^\chi$, they have to become degenerate as  
an unavoidable consequence of the (approximate) chiral symmetry in 
the strong interactions. The reason for this discrepancy may be related to 
the fact that in the heat bath the $\rho$ modifications as calculated in the 
present article are still hindered by  
the Goldstone nature of the pions. This observation does not
apply to $\rho$-interactions with nucleons in the finite 
density case. A complete understanding still requires further elucidation.

Resuming one of the  motivations of our analysis, we believe that the 
understanding of 
complex  strongly interacting systems that live in regions of density
and temperature far removed from equilibrium resides not only in
confronting theoretical calculations with experimental data, but also in
comparing the theories among themselves.

\acknowledgements 
We are grateful for productive conversations with 
 M. Wachs, J. Wambach and I. Zahed.
One of us (R. R.) acknowledges support
from the Alexander-von-Humboldt foundation as a Feodor-Lynen fellow
and thanks C. Gale for the hospitality during his visit to Montreal.  
C.G. is grateful to the Nuclear Theory Group of the State University of New
York at Stony Brook for its hospitality during part of a sabbatical
leave, when this work was initiated. 
This work is supported in part by the U.S. Department of Energy
under Grant No. DE-FG02-88ER40388, the Natural Sciences and
Engineering Research Council of Canada and the Fonds FCAR of
the Quebec Government.

\newpage

{\large TABLES} 

\begin{table}
\begin{tabular}{c|ccccc}
 $R$ & $I^GJ^P$ & $\Gamma_{tot}$ [MeV] & $\rho h$ Decay &
$\Gamma^0_{\rho h}$ [MeV] & $\Gamma^0_{\gamma h}$ [MeV] \\
\hline
$\qquad\omega(782)\qquad$ & $0^-1^-$ & 8.43 & $\rho\pi$ & $\sim 5$ & 0.72 \\
$h_1(1170)$   & $0^-1^+$ & $\sim 360$  & $\rho\pi$ & seen  &   ?  \\
$a_1(1260)$   & $1^-1^+$ & $\sim 400$  & $\rho\pi$ & dominant & 0.64 \\
$K_1(1270)$   & $\frac{1}{2}1^+$ & $\sim 90$ & $\rho K$  & $\sim 60$ &   ?  \\
$f_1(1285)$   & $0^+1^+$        & 25 & $\rho\rho$ & $\le$8   & 1.65  \\
$\pi'(1300)$  & $1^-0^-$        & $\sim 400$ & $\rho\pi$ & seen   & ?  \\
\end{tabular}
\caption{\it Mesonic Resonances $R$ with masses $m_R\le 1300$~MeV
and substantial branching ratios
into final states involving direct $\rho$'s (hadronic)
or $\rho$-like photons (radiative). }
\label{tabres}
\end{table}

\begin{table}
\begin{tabular}{c|ccccc}
 $R$ & $IF(\rho hR)$ & $G_{\rho hR}$ [GeV$^{-1}$] & $\Lambda_{\rho hR}$ [MeV] &
$\Gamma^0_{\rho h}$ [MeV] & $\Gamma^0_{\gamma h}$ [MeV] \\
\hline
$\qquad\omega(782)\qquad$ & 1 & 25.8  & 1000 & 3.5 & 0.72 \\
$h_1(1170)$               & 1 & 11.37 & 1000 & 300 & 0.60 \\
$a_1(1260)$               & 2 & 13.27 & 1000 & 400 & 0.66 \\
$K_1(1270)$               & 2 & 9.42  & 1000 &  60 & 0.32 \\
$f_1(1285)$               & 1 & 35.7  &  800 &   3 & 1.67  \\
$\pi'(1300)$               & 2 & 9.67  &  1000 & 300 & 0  \\
\end{tabular}
\caption{\it Results of our fit to the decay properties of $\rho$-$h$
induced mesonic resonances $R$ with masses $m_R\le 1300$~MeV (the
$f_1(1285)$ and $\pi'(1300)$ coupling constants are in units of
GeV$^{-2}$).}
\label{tabfit}
\end{table}
\end{document}